\documentclass[aps,prl,twocolumn,amsmath,amssymb,nofootinbib,
  superscriptaddress]{revtex4-2}

\usepackage{graphicx}
\usepackage{bm}
\usepackage[colorlinks]{hyperref}
\usepackage[usenames,svgnames]{xcolor}
\usepackage{makecell}
\usepackage[normalem]{ulem}

\newcommand{\zjlab}{Zhejiang Lab, Hangzhou, 311121, China}
\newcommand{\tsinghua}{Tsinghua University, Beijing, 100084, China}
\newcommand{\wuxi}{National Supercomputing Center in Wuxi, Wuxi, 214000, China}

\newcommand{\pcet}{National Research Center of Parallel Computer Engineering and Technology, Beijing 100190, China}
\newcommand{\ieu}{Information Engineering University, Zhengzhou, 450001, China}
\newcommand{\hnu}{Key Laboratory of Low-Dimensional Quantum Structures and Quantum Control of Ministry of Education, Department of Physics and Synergetic Innovation Center for Quantum Effects and Applications, Hunan Normal University, Changsha 410081, China
}

\begin{document}

\title{Verifying quantum supremacy experiments with multiple amplitude tensor network contraction}

\author{Yong Liu}
\thanks{These authors contribute equally to this work.}
\affiliation{\zjlab}

\author{Yaojian Chen}
\thanks{These authors contribute equally to this work.}
\affiliation{\tsinghua}

\author{Chu Guo}
\email{guochu604b@gmail.com}
\affiliation{\hnu}

\author{Jiawei  Song}
\affiliation{\wuxi}

\author{Xinmin Shi}
\affiliation{\ieu}

\author{Lin Gan}
\email{lingan@tsinghua.edu.cn}
\affiliation{\tsinghua}
\affiliation{\wuxi}

\author{Wenzhao Wu}
\affiliation{\wuxi}

\author{Wei Wu}
\affiliation{\wuxi}

\author{Haohuan Fu}
\email{haohuan@tsinghua.edu.cn}
\affiliation{\tsinghua}
\affiliation{\wuxi}

\author{Xin Liu}
\email{lucyliu_zj@163.com }
\affiliation{\zjlab}
\affiliation{\wuxi}

\author{Dexun Chen}
\affiliation{\wuxi}

\author{Zhifeng Zhao}
\affiliation{\zjlab}

\author{Guangwen Yang}
\affiliation{\zjlab}
\affiliation{\tsinghua}
\affiliation{\wuxi}

\author{Jiangang Gao}
\affiliation{\pcet}



\begin{abstract}
The quantum supremacy experiment, such as Google Sycamore [Nature \textbf{574}, 505 (2019)], poses great challenge for classical verification due to the exponentially-increasing compute cost. Using a new-generation Sunway supercomputer within $8.5$ days, we provide a direct verification by computing three million exact amplitudes for the experimentally generated bitstrings, obtaining an XEB fidelity of $0.191\%$ (the estimated value is $0.224\%$). The leap of simulation capability is built on a multiple-amplitude tensor network contraction algorithm which systematically exploits the ``classical advantage" (the inherent ``store-and-compute" operation mode of von Neumann machines) of current supercomputers, and a fused tensor network contraction algorithm which drastically increases the compute efficiency on heterogeneous architectures. Our method has a far-reaching impact in solving quantum many-body problems, statistical problems as well as combinatorial optimization problems.
\end{abstract}

\maketitle

%
%

Ever since initially visioned by Feynman in 1982~\cite{Feynman1982}, quantum computing has experienced $40$ years of theoretical and experimental developments~\cite{Shor1994,krantz2019quantum,huang2020superconducting,slussarenko2019photonic,blatt2012quantum,bruzewicz2019trapped,biamonte2017quantum,mcardle2020quantum}, starting to demonstrate a quantum advantage over classical computers in the era of noisy intermediate scale quantum computing~\cite{Preskill2018}. A major experimental milestone is the \textit{quantum supremacy} experiment conducted with the Google Sycamore $53$-qubit superconducting quantum processor in 2019~\cite{AruteMartinisQuantumSupremacy2019}, which demonstrates $10^9$ times better capability  for sampling a random quantum circuit (RQC) over the fastest classical supercomputer Summit at that time. The more recent $56$-qubit and $60$-qubit Zuchongzhi quantum processors are estimated to be around $26$ and $40,000$ times harder than Sycamore to classically simulate~\cite{WuPan2021,ZhuPan2021}. 

In the RQC sampling task, one runs a RQC on a (noisy) quantum processor and then measures it to produce a number of bitstrings (samples). While generating a number of samples is an easy task for quantum processors, simulating this task on a classical computer is a hard problem~\cite{BremnerDan2016,BoixoNeven2018,BoulandVazirani2019,HangleiterEisert2022}, even for noisy RQCs~\cite{AaronsonChen2017,AaronsonGunn2020} (noticing a recent work which proposed a polynomial but impractical algorithm for simulating constant-noise RQCs~\cite{AharonovVazirani2022}).
Several attempts have been made to narrow down the complexity gap set by Sycamore using the tensor network contraction (TNC) algorithm~\cite{MarkovShi2008},
powered by the recently developed excellent heuristic strategies to identify a near-optimal tensor network contraction order (TNCO)~\cite{GrayKourtis2020,HuangChen2021}.
Using a fused tensor contraction algorithm and a highly parallelized implementation on the new Sunway supercomputer, the runtime for computing a batch of correlated amplitudes for the depth-$20$ Sycamore RQC was reduced to about $300$ seconds~\cite{LiuChen2021}, which is currently further shortened to less than $150$ seconds by using a lifetime theory to reduce the slicing overhead and increase the compute density~\cite{ChenYang2022}. For computing uncorrelated amplitudes, a recursive multi-tensor contraction algorithm is recently proposed and used to compute millions of amplitudes for Sycamore RQCs up to depth $16$~\cite{KalachevYung2021}. However validating the depth-$20$ case by exactly computing a large number of uncorrelated amplitudes is still out of reach.
To this end we note that to attack the claim of quantum supremacy, several works have directly simulated the noisy RQC sampling by exploring biased noises to drastically reduce the computational cost~\cite{PanZhang2021b,KalachevYung2021b,GaoChoi2021}. 
Here we focus on computing exact amplitudes instead, so as to provide a verification to noisy RQCs.

In this work, we manage to, for the first time, compute three million uncorrelated amplitudes of the most complicated depth-$20$ Sycamore RQC (referred to as Sycamore afterwards), using $107,520$ SW26010P CPUs ($41,932,800$ cores) for $8.5$ days. 
Our simulation efficiency is at least three orders of magnitude faster than the best existing records which have successfully computed exact amplitudes of Sycamore~\cite{LiuChen2021,PanZhang2021,ChenYang2022} (computing a batch of correlated amplitudes will only induce a small computational overhead compared to computing a single amplitude for TNC algorithm~\cite{VillalongaMandra2018,HuangChen2021}), and is only about $2.5$ times slower per amplitude compared to the Sycamore quantum processor itself.
The jump of simulation capability is made possible by mainly two algorithms, one focus more on the algorithmic side and the other focus more on the implementation side. 
On the algorithmic side, we systematically explore the ``classical advantage" of storing and reusing intermediate tensor results, which theoretically lowers the computation cost of computing millions of uncorrelated amplitudes of Sycamore by at least three orders of magnitude. 
On the implementation side, we build our simulator with a fused tensor network contraction algorithm to largely reduce data movement and increase compute density, and an adaptive parallelization scheme that fully utilizes the hundreds of cores in each processor for different sizes of tensors, which enables us to almost fully achieve the theoretical speedup.

Our results provide a concrete verification for Sycamore, which is more than $10^3$ times harder than simulating the noise sampling task itself. With further improvements for larger circuit sizes, we vision that the Zuchongzhi series of RQCs are also verifiable in the near term. 
Although we have focused on computing uncorrelated amplitudes of a RQC, our method is completely general for contracting a large number of similar tensor networks (TNs) which share a significant portion of common tensors. Therefore the ma-TNC algorithm and the parallelization techniques developed in this work could have a far-reaching impact beyond simulating RQCs, such as solving quantum many-body problems~\cite{Orus2019}, statistics problems~\cite{MezardVirasoro1987} or combinatorial optimization problems~\cite{MezardZecchina2002,MezardMontanari2009}, which can generally be formulated as tensor network contraction problems~\cite{GarciasaezLatorre2012,GuoArad2023}. 

\begin{figure*}
\includegraphics[width=2\columnwidth]{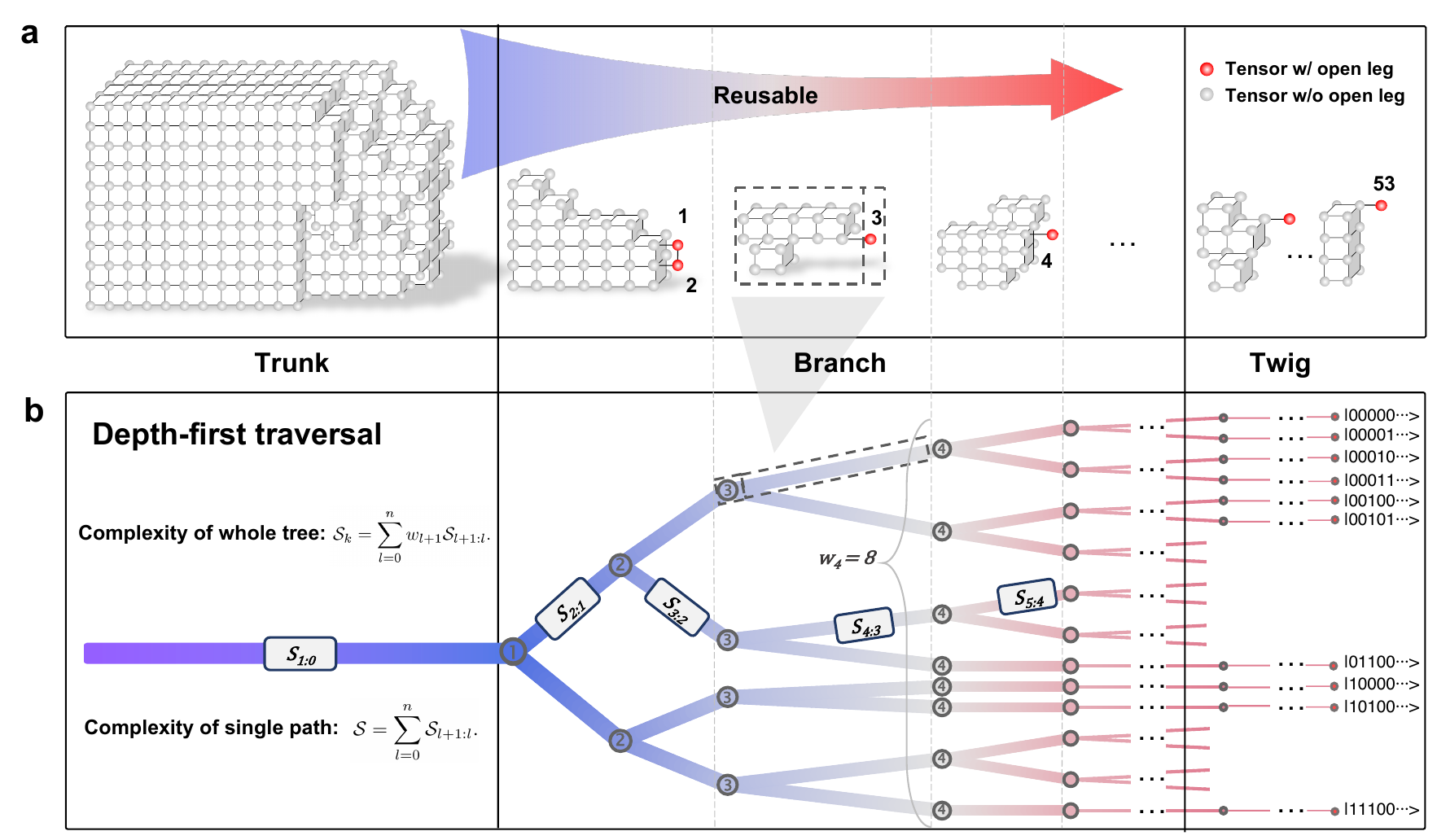}
\caption{Demonstration of the ma-TNC algorithm. (a) Division of the original tensor network formed in the TNC algorithm into consecutive sub tensor networks, where each sub tensor network starts with a bright tensor containing one or more output indices and ends before the next bright tensor. (b) Organizing the contraction of the $k$ tensor networks resulting from computing $k$ amplitudes into a tree, where each node of the tree corresponds to an output index and the edge after it corresponds to a specific choice of this index. A full path along the tree from left to right corresponds to contracting one tensor network and traversing the tree means to contract all the tensor networks. Generally for computing a large number of uncorrelated amplitudes, the left most edge (the trunk) contains most of the tensors, the number of edges in a vertical layer grows exponentially in the central part (the branch) and stops growing in the tail (the twig).
}
\label{fig:fig1}
\end{figure*}

\textbf{Multiple-amplitude simulation with static and optimal tensor reuse.} For a quantum computer, due to the no-cloning theorem~\cite{WoottersZurek1982}, the cost of generating $k$ samples from a RQC, denoted as $\mathcal{S}_k$, is strictly linear against $k$, namely $\mathcal{S}_k = k \mathcal{S}$ with $\mathcal{S}$ the cost of producing a single sample. 
When simulating RQC on classical computers, the relation between $\mathcal{S}_k$ and $\mathcal{S}$ depends on the specific method to use. In the past three years, the method of choice to simulate Sycamore(-like) quantum processors has gradually converged to the TNC algorithm due to its relatively low computational cost and well-controlled memory usage by using the slicing technique~\cite{GrayKourtis2020,HuangChen2021}. The TNC algorithm transforms the whole quantum circuit into a large TN (the original TN) with $n$ uncontracted tensor indices (the output indices), each corresponding to an output qubit state.
Computing the amplitude of a given bitstring amounts to selecting particular elements of the uncontracted tensor indices, resulting in a TN with no uncontracted indices. Computing $k$ amplitudes will result in $k$ TNs, which only differ in the choice of the output indices of the original TN.

Existing approaches using TNC mostly compute a single amplitude or a correlated batch of amplitudes (referred to as sa-TNC in the following) one time~\cite{GuoWu2019,GuoHuang2021,HuangChen2021,PanZhang2021}, while our ma-TNC algorithm computes $k$ uncorrelated amplitudes in a single run, which proceeds as follows.
We first assume that a TNCO for computing one amplitude has been obtained. 
We refer to those tensors in the original TN which contain at least one output index as the \textit{bright tensors}. Following the TNCO, whenever a bright tensor is met, there could be a \textit{branching}, which means that several TNs among all the $k$ TNs share the same tensors till this bright tensor. Therefore, the computations before contracting this bright tensor can be perfectly reused among them.
To systematically identify all such reusable patterns, we divide the original TN into many sub tensor networks (blocks) along the TNCO, where each block starts from a bright tensor and ends before the next bright tensor (the first block has no bright tensor), as shown in Fig.~\ref{fig:fig1}(a).
Then we organize the $k$ TNs into a tree (the reuse tree) as shown in Fig~.\ref{fig:fig1}(b). Each node of the tree corresponds to one output index, while the edge after the node corresponds to a particular choice of this index. The nodes in the same vertical line correspond to the same output index and form a \textit{layer}. A bright tensor containing multiple output indices would correspond to multiple layers. There could also exist tensors between successive bright tensors along the TNCO that do not contain any output indices, which are assumed to live on the edges.
Given these correspondences, a full path from left to right along the tree corresponds to contracting the TN for computing one amplitude, and traversing the tree corresponds to contracting all the $k$ TNs for $k$ amplitudes. It is clear that all the intermediate tensors before a certain node of the tree can be reused for all the subpaths after (and including) this node, thus an optimal reuse strategy would be to reuse all such tensors to reduce the computational cost.

We denote the computational cost of one edge connecting a node in the $l$-th layer and another node in the $l+1$-th layer as $\mathcal{S}_{l+1,l}$, then the cost of one path is $\mathcal{S} = \sum_{l=0}^n\mathcal{S}_{l+1:l}$.
Denoting the number of nodes in the $l$-th layer as $w_l$ (the \textit{width}), which is the same as the number of edges between the $(l-1)$-th and $l$-th layers, then the computational cost between the $l$-th layer and the $(l+1)$-th layer is $w_{l+1} \mathcal{S}_{l+1:l}$ for optimal reuse, and the total cost is 
\begin{equation}\label{eq:fulcpx}
\mathcal{S}_k = \sum_{l=0}^{n} w_{l+1} \mathcal{S}_{l+1:l}.
\end{equation}
$w_l$ is non-decreasing with $l$ which satisfies $w_1=1$, $w_{n+1}=k$ and $1\leq w_{l}\leq k$ for $1 < l \leq n$. 
For computing a large number (but not exponentially large) of uncorrelated amplitudes, $w_l$ will typically grow exponentially at the beginning before it saturates. 
We can see that that $\mathcal{S}_k < k\mathcal{S}$ in general. Therefore when searching for an optimal TNCO, we choose to directly minimize $\mathcal{S}_k$ instead of $\mathcal{S}$ (for optimization we have used $\mathcal{S}_k$ as the loss function in the KaHyPar package~\cite{kahypar}). For Sycamore, we observe that this choice can easily lower the overall computational cost by more than $10$x (Details can be found in SM~\cite{supp} Sec.~IV).

To minimize the memory cost, one can perform a \textit{depth-first traversal} of the tree, where one only needs to store all the intermediate tensors at the nodes in the branch along a single path from left to right. For Sycamore we found that the amount of memory required for a reuse-oriented computing of $3M$ amplitudes is only about two times that of computing a single amplitude. In comparison in the breath-first traversal one needs to store all the intermediate tensors at one layer (scales with $k$), in which the memory usage could easily explode.

We summarize the defining features of our ma-TNC algorithm: 1) it directly minimizes the multi-amplitude cost in Eq.(\ref{eq:fulcpx}) when searching for a near optimal TNCO and 2) it organizes the computation into a static tree and performs a depth-first traversal of the tree to accomplish the computation, which achieves optimal reuse of intermediate computations with minimal memory cost for a given TNCO. The static nature of our algorithm makes the tensor contraction pattern and the memory allocation predetermined, which is extremely important for massive parallelization.
To this end we stress that the idea of computing multiple uncorrelated amplitudes simultaneously to reduce redundant intermediate computation has already been explored in Ref.~\cite{KalachevYung2021}, where it is estimated that computing $3M$ uncorrelated amplitudes could be done using Summit within $7.5$ days. However, our approach is very different from Ref.~\cite{KalachevYung2021}. We formulate the whole computation as a static reuse tree for a given TNCO, as such the memory and computational cost, as well as the whole parallelization strategy are completely determined before we actually perform the calculations (since the computational cost is known for each given TNCO, we also use it as the loss function to optimize the TNCO). In comparison, Ref.~\cite{KalachevYung2021} uses a dynamical global cache whose entries are frequently inserted and deleted and the reusable intermediate tensors are only determined during the actual calculations. For large scale RQCs, the latter approach is likely to affect the parallelization efficiency and could easily run into memory issue since the memory cost is not predetermined. 

\textbf{Parallelizing the ma-TNC method over 40 million cores of the new Sunway supercomputer.}
In our large-scale implementation on the new Sunway supercomputer, we use a two-level parallelization scheme. In the first level we use the slicing technique as a standard practice for the TNC algorithm to produce $2^{22}$ ($\approx 4$ million) slices for parallel processing over CPUs~\cite{HuangChen2021,LiuChen2021}. In the second level we contract each slice using ma-TNC on each CPU (which contains $384$ cores).
For computing $k=3M$ amplitudes, we found a TNCO for which the ideal speedup compared to sa-TNC is $1328$x, while the actual speedup using swTT (our previous tensor contraction implementation~\cite{LiuChen2021}) is only $40$x. The reason for this slow down is that for computing a large number of uncorrelated amplitudes, the calculation is dominated by very small tensor contractions with low compute density.
To restore the computational efficiency we propose a fused TNC algorithm, combined with an adaptive parallelization scheme that works differently with different tensor sizes. The central goal of the fused TNC algorithm is to perform several successive tensor contractions together so as to reduce data movement.
With these techniques we are able to restore the speedup of ma-TNC against sa-TNC from $40$x to around $1248$x. Details of the fused TNC algorithm can be found in SM~\cite{supp} Sec.~V.

\begin{figure}
\center
\includegraphics[width=\columnwidth]{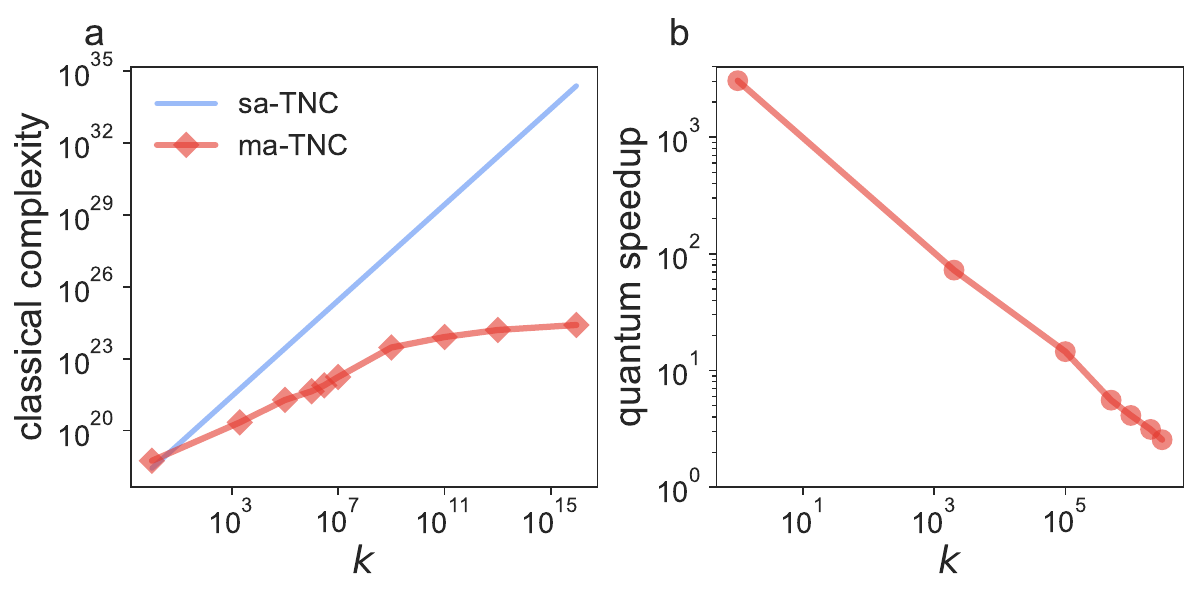}
\caption{(a) The red line with diamond shows the scaling of the theoretical complexity of our ma-TNC against the number of amplitudes $k$ for Sycamore, while the blue line shows the linear scaling of the sa-TNC as a reference. (b) The quantum speedup of Sycamore against our ma-TNC, defined as our classical runtime divided by the quantum runtime, as a function of $k$. 
}
\label{fig:fig3}
\end{figure}

\textbf{Verification of Sycamore.}
We first evaluate the theoretical computational cost and actual performance of our ma-TNC for simulating Sycamore. 
In Fig.~\ref{fig:fig3}(a) we show the scaling of the theoretical cost of ma-TNC against $k$, based on an optimal TNCO found by minimizing Eq.(\ref{eq:fulcpx}). The scaling of sa-TNC is shown as a reference. We can clearly see that the cost of ma-TNC scales only sublinearly against $k$. For $k\approx 10^6$, the cost of ma-TNC is already lower than sa-TNC by more than three orders of magnitude. 

In Fig.~\ref{fig:fig3}(b) we show the actual performance of our ma-TNC using our well-optimized implementation on the new Sunway supercomputer, where we have also used the quantum runtime of Sycamore as the benchmarking baseline. The runtime for computing a single amplitude is assumed to be equivalent to that for generating a perfect sample using the TNC algorithm, since one could easily adjust the TNC algorithm to compute a small batch of correlated amplitudes with negligible overhead, and obtain a perfect sample with unit probability from the batch~\cite{VillalongaMandra2018,HuangChen2021}. Since we only compute exact amplitudes (perfect samples), the complexity of generating $k$ perfect samples is assumed to be equivalent to that of generating $k/f$ noisy samples with fidelity $f$~\cite{MarkovBoixo2018} (therefore the task of computing $3M$ exact uncorrelated amplitudes is $1500$x times harder than generating $1M$ noisy samples with $f=0.2\%$). 
The quantum speedup is then defined by the classical runtime of ma-TNC divided by the quantum runtime of Sycamore multiplied by $1/f$. 
We can see that while the quantum speedup is more than $3,000$x for $k=1$, it drastically decreases to $4$x for $k=1M$ and $2.5$x for $k=3M$. 

For completeness, we list in Table.~\ref{tab:speedup} the ideal and actual speedups of ma-TNC over sa-TNC for computing $1M$ uncorrelated amplitudes for Sycamore RQCs of \textit{different depths}. As comparison, the speedup reported in Ref.~\cite{KalachevYung2021} is $10000$x for depth $16$ for $2M$ amplitudes, and the estimated actual speedups for depths $18$ and $20$ are $5193$x and $1022$x for $2.5M$ and $3M$ amplitudes respectively. Taking into account that the speedup is more significant with more amplitudes, our ideal speedup (where the ma-TNC is assumed to be implemented with the same efficiency as sa-TNC), as well as our actual speedups for depths $18$ and $20$, are generally higher than Ref.~\cite{KalachevYung2021} (the actual speedup for depth $16$ is significantly lower than the ideal speedup, which is because that our implementation is better tuned for deeper circuits). 

\begin{table}[!htb]
  \begin{center}
    \caption{The ideal and actual speedups of ma-TNC over sa-TNC for computing one million uncorrelated amplitudes. The first column lists the Sycamore RQCs with different depths. The columns ``FPOs (s)'' and ``FPOs (m)'' are the number of floating point operations for sa-TNC and ma-TNC respectively. The maximum size of intermediate tensors is set to be $2^{31}$ for all cases. Here the actual speedups are estimated by calculating a single slice on one CPU for both algorithms.}
    \label{tab:speedup}
    \begin{tabular}{c|c|c|c|c}
    \hline
    \hline 
    Depth & FPOs (s) &  FPOs (m) & Ideal speedup & Actual speedup  \\
    \hline 
    $16$ & $1.3\times 10^{17}$   & $1.4\times 10^{19}$ & $9286$ & $2311$  \\
    $18$ & $5.3\times 10^{17}$   & $5.0\times 10^{19}$ & $10600$ & $5393$  \\
    $20$ & $5.5\times 10^{18}$   & $5.1\times 10^{21}$ & $1080$ & $737$  \\
\hline
    \end{tabular}
  \end{center}
\end{table}

\begin{figure}
\center
\includegraphics[width=0.8\columnwidth]{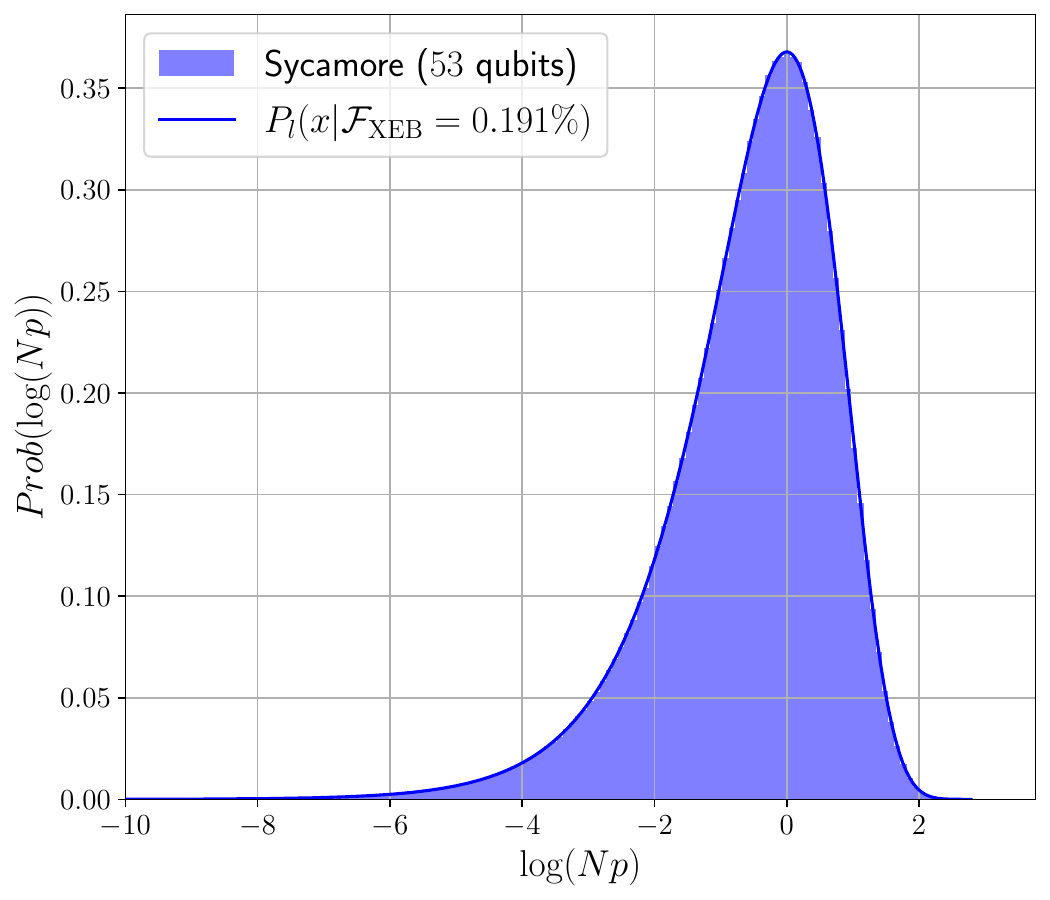}
\caption{ Histogram for the distribution of the probabilities of the three million experimentally generated bitstrings from Sycamore, where the x-axis is log-rescaled. The blue solid line denotes the corresponding theoretically prediction under the same XEB fidelity as defined in Eq.(\ref{eq:fxeb}).
}
\label{fig:fig4}
\end{figure}

We also directly compute the exact amplitudes of $3M$ experimentally generated samples by Sycamore, which is done by using $107,520$ SW26010P CPUs for $8.5$ days ($203$ hours). Our results show that the exact XEB fidelity for these bitstrings is $\mathcal{F}_{{\rm XEB}} = (0.191\pm 0.058)\%$, which closely matches the estimated value of $(0.224\pm 0.021)\%$. We plot in Fig.~\ref{fig:fig4} the histogram of the obtained amplitudes and compare them to the theoretical probability density function for the rescaled bitstring probability $Np$ ($N=2^n$ and $p$ is the probability) under the same XEB fidelity, defined as
\begin{equation}\label{eq:fxeb}
P_l(x|\mathcal{F}_{{\rm XEB}}) = \left(\mathcal{F}_{{\rm XEB}} x + (1-\mathcal{F}_{{\rm XEB}}) \right) e^{-x},
\end{equation}
with $x=Np$. We can see that they agree well with each other, which means that the bitstrings generated by Sycamore indeed obeys the Porter-Thomas distribution with the estimated XEB fidelity. Our results thus provide a strong consistency check for the Sycamore quantum supremacy experiment.

\textbf{Discussions.} The Sycamore quantum processor could generate one million noisy samples with $0.2\%$ fidelity in $200$ seconds. In comparison we have computed three million amplitudes on the new Sunway supercomputer within $8.5$ days, a task that is more than $10^3$ harder than that performed by Sycamore.
Taking into account that the complexity increases by the Zuchongzhi series quantum processors are not dramatic compared to Sycamore (mostly due to that the increase is mostly in terms of the number of qubits instead of the gate fidelities~\cite{ZlokapaLidar2023}), we vision that those quantum processors can also be simulated in near term. In the meantime, we mention that the cost of our calculation is around $1.5$ million Chinese Yuan, which is probably more expensive than the experiments performed on Sycamore. After we have finished this work, we notice the newest RQC sampling task performed on a quantum processor with $70$ qubits and $24$ depths~\cite{Google2023b}, which is likely beyond our reach.

Other than simulating RQCs, our results also represent a major jump of the ability in contracting a large number of tensor networks with the same structure and sharing most of the tensors in common, which is a very universal situation that could be encountered in computational physics and combinatorial optimization problems and thus could be of very wide interest.

\begin{acknowledgements}
We thank Xun Gao, Man-Hong Yung, Xiaobo Zhu, Zuoning Chen for helpful discussions and comments. 
This research was supported in part by the National Key Research and Development Plan of China (Grant No. 2020YFB0204800), National Natural Science Foundation of China (Grant No. T2125006, U1839206), Jiangsu Innovation Capacity Building Program (Project No. BM2022028).
C. G acknowledges support from National Natural Science Foundation of China under Grants No. 11805279.
The three million experimentally generated bitstrings are downloaded from https://doi.org/10.1038/s41586-019-1666-5. 
The bitstrings together with the calculated amplitudes are available at https:/github.com/leao077/ma\_TNC.
\end{acknowledgements}

\bibliographystyle{apsrev4-2}
%

\end{document}


\title{Supplementary Information: Verifying quantum supremacy experiments with multiple amplitude tensor network contraction}


\date{\today}

\maketitle

\tableofcontents

\section{Random quantum circuits}

\begin{figure}
\includegraphics[width=\columnwidth]{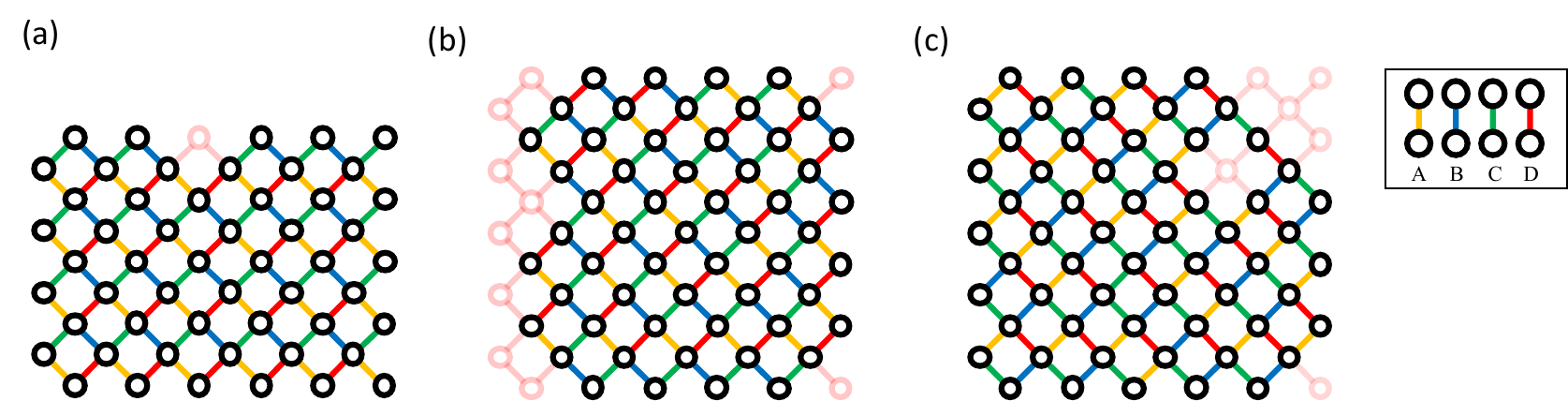}
\caption{Patterns of two-qubit gate layers for (a) Sycamore ($53$ qubits), (b) Zuchongzhi-$2.0$ ($56$ qubits) and (c) Zuchongzhi-$2.1$ ($60$ qubits).
}
\label{fig:figS1}
\end{figure}

A random quantum circuit (RQC) often consists of interlacing layers of single-qubit gates and two-qubit gates. Each layer of two-qubit gates is counted as one cycle (depth), with a geometry that is friendly for the underlying quantum hardware.
The Sycamore~\cite{AruteMartinisQuantumSupremacy2019} and Zuchongzhi series~\cite{WuPan2021,ZhuPan2021} of quantum processors all share the same geometrical structure but differ in the total number of qubits.
For the RQC sampling tasks on these quantum processors, the single-qubit gates are randomly and independently selected from the set $\{\sqrt{X}, \sqrt{Y}, \sqrt{W}\}$, which are defined as
\begin{align}
\sqrt{X} &= \frac{\sqrt{2}}{2} \left[ \begin{array}{cc} 1 & -\im \\ -\im & 1 \end{array} \right]; \\ 
\sqrt{Y} &= \frac{\sqrt{2}}{2} \left[ \begin{array}{cc} 1 & -1 \\ 1 & 1 \end{array} \right]; \\ 
\sqrt{W} &= \frac{\sqrt{2}}{2} \left[ \begin{array}{cc} 1 & -\frac{\sqrt{2}}{2}(1+\im) \\ \frac{\sqrt{2}}{2}(1-\im) & 1 \end{array} \right],
\end{align}
and the two-qubit gate is the $5$-parameter fSim gate 
\begin{align}
\left[\begin{array}{cccc} 1 & 0 & 0 & 0 \\ 0 & e^{\im(\Delta_+ + \Delta_-)}\cos(\theta) & -\im e^{\im(\Delta_+ - \Delta_{-,{\rm off}})}\sin(\theta) & 0 \\ 0 & -\im e^{\im(\Delta_+ + \Delta_{-,{\rm off}})}\sin(\theta) & e^{\im(\Delta_+ + \Delta_-)}\cos(\theta) & 0 \\ 0 & 0 & 0 & e^{\im(2\Delta_+ - \phi)} \end{array} \right]. \nonumber
\end{align}
The values of those parameters are obtained through certain calibration procedures.

In all the three quantum supremacy experiments, $4$ different patterns of two-qubit layers are used, which are denoted as $A,B,C,D$. Those $4$ patterns are shown in Fig.~\ref{fig:figS1}, where each empty circle denotes one qubit and  each line between two circles denotes a two-qubit gate operation. The shaded qubits and lines mean that those qubits as well as gate operations are not used in experiments, possibly due to high error rates. As a result each pair of nearest-neighbour qubits is operated on by a two-qubit gate once and only once in each $4$ cycles containing all $A,B,C,D$. A RQC of a fixed depth is then formed by repeating these $4$ patterns, with randomly generated single-qubit gate layers in between. For example a RQC of depth $8$ is chosen as $ABCDCDAB$, and a RQC of depth $12$ is $ABCDCDABABCD$. 
The overall design principle of RQC is to make the classical simulation cost as high as possible given the same number of qubits and the same amount of quantum gate operations.

\section{Tensor network contraction algorithm}

\begin{figure}[htbp]
    \centerline{\includegraphics[width=0.5\textwidth]{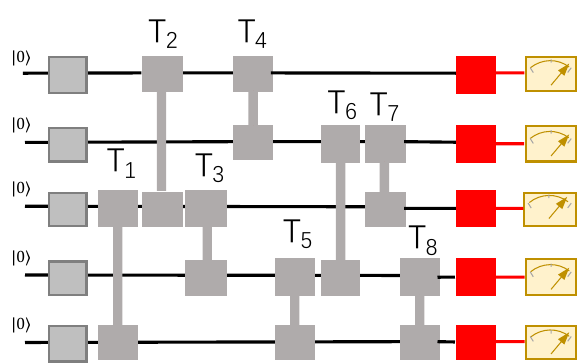}}
    \caption{Demonstration of a $5$-qubit quantum circuit. On the left hand side is the input quantum state of the quantum circuit which is chosen as $\vert 0\rangle$ for all the qubits, while on the right hand side of the quantum circuit a quantum measurement is performed to obtain a bitstring. The legs marked red on the right hand side correspond to the indices of the output quantum state $\vert \psi\rangle$.}
    \label{fig:figS2}
\end{figure}

Generally, a quantum circuit generates a final quantum state $\vert \psi\rangle$ by applying a series of quantum gate operations onto an initial state, which is often chosen as $\vert 0^{\otimes n}\rangle$ for an $n$-qubit system. 
The quantum state $\vert \psi\rangle$ can be written as
\begin{align}
\vert \psi\rangle = \sum_{\sigma_1, \sigma_2, \dots, \sigma_n} c_{\sigma_1, \sigma_2, \dots, \sigma_n} \vert \sigma_1, \sigma_2, \dots, \sigma_n\rangle,
\end{align}
where $\sigma_j \in \{0, 1\}$, $\vert \sigma_1, \sigma_2, \dots, \sigma_n\rangle$ denotes a particular computational basis, and $c_{\sigma_1, \sigma_2, \dots, \sigma_n}$ is the amplitude for this basis, which satisfies
\begin{align}
c_{\sigma_1, \sigma_2, \dots, \sigma_n} = \langle \sigma_1, \sigma_2, \dots, \sigma_n \vert \psi\rangle. 
\end{align}
$|c_{\sigma_1, \sigma_2, \dots, \sigma_n}|^2$ is the probability of obtaining the bitstring $\sigma_1 \sigma_2 \dots \sigma_n$ if one performs a quantum measurement at the end of the quantum circuit. A simple $5$-qubit quantum circuit is demonstrated in Fig.~\ref{fig:figS2}, where the $5$ legs that are marked red correspond to the indices of the rank-$5$ coefficient tensor $c_{\sigma_1, \sigma_2, \dots, \sigma_n}$ (the output indices) of the final quantum state $\vert \psi\rangle$.

\begin{figure}[htbp]
    \centerline{\includegraphics[width=0.5\textwidth]{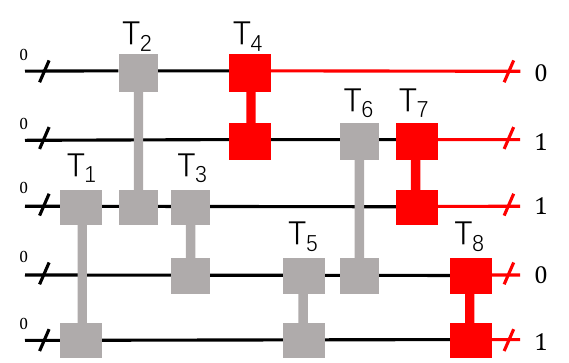}}
    \caption{The tensor network corresponding to the quantum circuit in Fig.~\ref{fig:figS2}, where the single-qubit gates have been absorbed into the two-qubit gates using gate fusion. The black dashes on the indices at the left boundary of the tensor network mean to take the slice $0$ for those indices which correspond to the qubit state $\vert 0\rangle$, while the red dashes on the indices at the right boundary of the tensor network mean to take the particular slice according to the bitstring for which the amplitude is to be computed (the bitstring $01101$ is shown as a specific example).}
    \label{fig:figS3}
\end{figure}

A quantum circuit can be naturally mapped into a tensor network (TN): a single-qubit state ($\vert 0\rangle$ or $\vert 1\rangle$) is a rank-$1$ tensor, a single-qubit gate is a rank-$2$ tensor and a two-qubit gate is a rank-$4$ tensor. Taking the quantum circuit in Fig.~\ref{fig:figS2} as an example, it can be mapped into the TN shown in Fig.~\ref{fig:figS3} (The rank-$1$ and rank-$2$ tensors are pre-absorbed into the rank-$4$ tensors, which results in a simplified TN containing only rank-$4$ tensors.). There are $5$ uncontracted indices in the TN in Fig.~\ref{fig:figS3}, which are the output indices in Fig.~\ref{fig:figS2}. Contracting this TN, one would obtain the coefficient tensor $c_{\sigma_1, \sigma_2, \dots, \sigma_n}$, namely all the amplitudes. 
If one selects a particular slice $b_j$ for each output index $j$, the resulting TN would contain no uncontracted indices, and contracting this TN one would obtain the amplitude $\langle \bold{b}\vert \psi\rangle$ for the bitstring $\bold{b}=b_1b_2\dots b_n$. 
Computing a batch of correlated amplitudes means to select particular slices for most of the output indices, with the rest left open~\cite{HuangChen2021,PanZhang2021,LiuChen2021}.

The computational cost of contracting a TN is crucially dependent on a good tensor network contraction order (TNCO). However, finding the optimal TNCO for an arbitrary TN is an NP-hard problem and is almost impossible for our case with hundreds of tensors. Important progresses have been made in the last three years to heuristically find a near optimal TNCO~\cite{GrayKourtis2020,HuangChen2021}, which significantly reduce the computational cost of computing a single amplitude or a batch of correlated amplitudes. The most difficult quantum supremacy experiments performed till now are listed in Table.~\ref{tab:tab1}, where we have also shown the computational cost for computing a single amplitude (the data is taken from Ref.~\cite{ZhuPan2021}). 
By making the equivalence between the cost of computing one exact amplitude (or generating one perfect sample) and the cost of generating a noisy sample with XEB fidelity $f$ times $1/f$~\cite{MarkovBoixo2018}, and taking the fastest classical runtime reported for computing a correlated batch of amplitudes of Sycamore~\cite{ChenYang2022} as the baseline for classical computers, we can see that the quantum speedup, defined as the classical runtime $t_{\CPU}$ divided by the quantum runtime $t_{\QPU}$:
\begin{align}
\speedup = \frac{t_{\CPU}}{t_{\QPU}},
\end{align}
is $\speedup \approx 1.7\times 10^3$ for Sycamore (the quantum speedup is initially estimated as $10^9$x in Ref.~\cite{AruteMartinisQuantumSupremacy2019}), $\speedup \approx 4.4\times 10^4$ for Zuchongzhi-$2.0$ and $\speedup \approx 6.7\times 10^7$ for Zuchongzhi-$2.1$. We can also see that the effective speedups of Zuchongzhi-$2.0$ and Zuchongzhi-$2.1$ compared to Sycamore are $26$x and $4\times 10^4$x respectively.

\begin{table}[!htb]
\centering
\caption{The most difficult random quantum circuit sampling tasks performed on Sycamore and the Zuchongzhi series quantum processors. The circuit size (Circuit), XEB fidelity (Fidelity), number of measured bitstrings (k) and the corresponding runtime on the quantum processor, as well as the classical computational cost for computing a single (exact) amplitude are listed. The classical computational cost for a single amplitude is computed using an optimal tensor network contraction order found by the package cotengra~\cite{GrayKourtis2020}.
}
\label{tab:tab1}
\begin{tabular}{ccccccc}
\hline
\hline
Processor & Circuit & Fidelity & k & Runtime (QPU) &  Classical computational cost (single-amplitude)  \\
\hline
Sycamore & $53$ qubit $20$ depth & $0.224\%$ & $3\times 10^6$ & $600$ s & $1.63 \times 10^{18}$ \\
Zuchongzhi-$2.0$ & $56$ qubit $20$ depth  & $0.0662\%$ & $1.9\times 10^7$ & $1.2$ h &  $1.65 \times 10^{20}$ \\
Zuchongzhi-$2.1$& $60$ qubit $24$ depth & $0.0336\%$ & $7\times 10^7$ & $4.2$ h &  $4.68 \times 10^{23}$  \\
\hline
\end{tabular}
\end{table}

\section{Tree Based Reuse Strategy}
\subsection{Reusability between amplitudes}
The tensor network contraction (TNC) algorithm converts the problem of computing one amplitude into the problem of contracting a TN with no uncontracted indices. For computing $k$ amplitudes there are $k$ TNs to be contracted. Naively, one could contract those TNs one by one independently (referred to as the sa-TNC algorithm), for which the complexity $\fullcpx$ for computing $k$ amplitude is related to the cost $\singlecpx$ of computing one amplitude as
\begin{align}\label{eq:fullcpx}
\fullcpx = k \singlecpx.
\end{align}
Eq.(\ref{eq:fullcpx}) is also the way that the classical cost is evaluated in Ref.~\cite{WuPan2021,ZhuPan2021}. 
However, the $k$ TNs formed for computing $k$ amplitudes are not completely different, instead they have exactly the same geometrical structure and moreover, a major portion of tensors are the same for all those TNs. For the depth-$20$ Sycamore RQC, there are about $400$ tensors in each TN, and about $350$ among them are the same for all the $k$ TNs. Therefore it is possible to reuse the intermediate computations to largely reduce the cost of computing $k$ amplitudes, namely $\fullcpx$.

Noticing that the tensors which are different among the $k$ TNs (namely the bright tensors as defined in the main text) are generally located at the tails of the TNCO, therefore a simple approach to reuse the intermediate computations is to store the intermediate tensor which is immediately before the first bright tensor (which is trunk as defined in the main text). However, with this global approach one is only able to reduce $\fullcpx$ by several times at most, and the reduction will decrease as $k$ increases. Better reuse is possible by realizing that if one takes out the TNs for a few amplitudes, the common tensors shared by them could be more than $350$, therefore a larger ``trunk'' for these TNs. An optimal reuse scheme would thus need to make use of all such microscopic patterns to reduce the computational cost, which can be systematically achieved using our multi-amplitude contraction tree (simply referred to as the reuse tree afterward) as follows.



\subsection{Structure of the reuse tree}
\begin{figure}[htbp]
    \centerline{\includegraphics[width=0.45\textwidth]{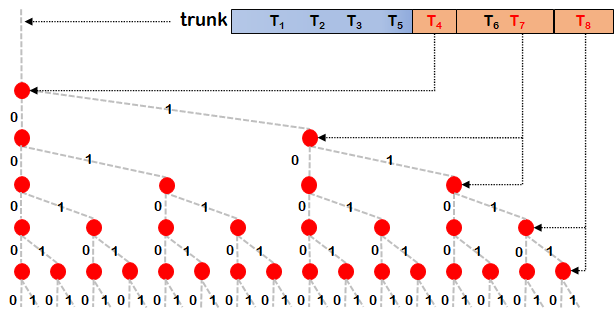}}
    \caption{The multi-amplitude contraction tree for computing all the amplitudes of the quantum circuit in Fig.~\ref{fig:figS2}, with a tensor network contraction order $T_1T_2T_3T_5T_4T_6T_7T_8$. }
    \label{fig:tree}
\end{figure}

Here we build a concrete reuse tree for the quantum circuit in Fig.~\ref{fig:figS2}, assuming that all the amplitudes are computed ($k=2^5=32$). The problem is first mapped to the contraction of the tensor network in Fig.~\ref{fig:figS3}, along the TNCO $T_1T_2T_3T_5T_4T_6T_7T_8$. The reuse tree for computing all the $32$ amplitudes is shown in Fig.~\ref{fig:tree} (the tree in the main text is from left to right while the tree presented here is from top down). We can see that the tensors $T_1T_2T_3T_5$ forms the trunk and the rest forms the branch. There is no twig since we compute all the amplitudes for which there will always be branching whenever a bright tensor is met, namely the width $w_l$ of the $l$-th layer of nodes grows exponentially as $w_l=2^{l-1}$. The bright tensors $T_7$ and $T_8$ contains two output indices each and thus each of them correspond to two successive layers. The tensor $T_6$ between the bright tensors $T_4$ and $T_7$ contains no output indices and thus lives on the edges between the first and second layers.

\subsection{Reuse tree traversal strategy with the lowest memory cost}

\begin{figure}[htbp]
    \centerline{\includegraphics[width=0.45\textwidth]{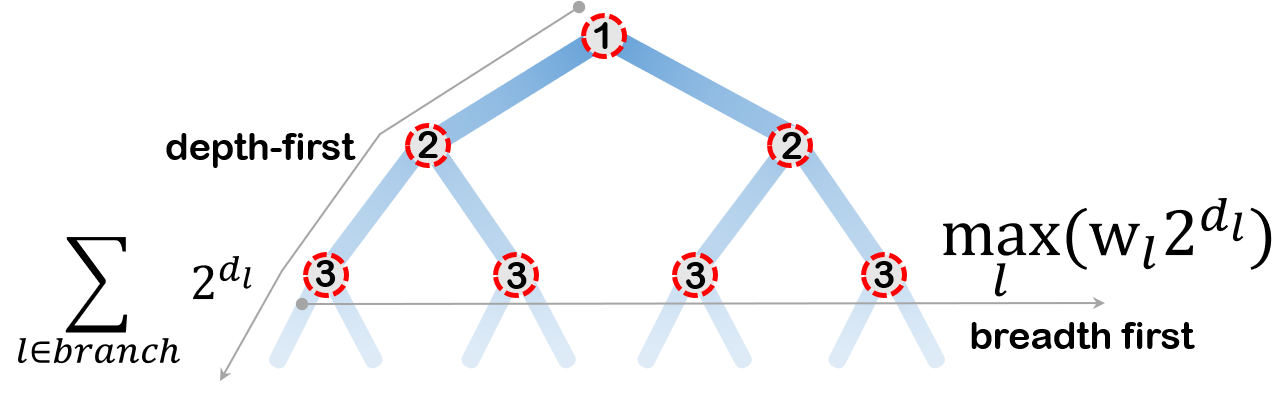}}
    \caption{A comparison of the memory costs for BFT and DFT. In BFT the memory cost is determined by the largest memory of all the intermediate tensors on one layer, while in DFT the memory cost is determined by all the intermediate tensors on a single path from root to a leaf.}
    \label{fig:traversal}
\end{figure}

After building up the reuse tree, it is clear that all the possible reusable points are located at the nodes of the tree. More concretely the intermediate tensor immediately before a node can be reused by all the subpaths grown from this node. Therefore 
the optimal $\fullcpx$ as shown in the main text can be achieved as long as the tree is traversed and all those intermediate tensors are properly reused.
However, the memory cost depends on the traversal strategy. A trivial reuse strategy is to store all those intermediate tensors and reuse them when required, which has the largest memory cost.

Typically to balance between computation and memory cost, one would choose an upper bound for the largest size of the intermediate tensors (such as $2^{30}$ which is $8$ GB for single precision) during the first-level slicing. To compute a single amplitude only one intermediate tensor needs to be stored at least. While for optimal reuse one often has to store several copies of the intermediate tensor and memory would be one of the major issues to be considered when designing a multi-amplitude TNC algorithm.

To reuse the intermediate computations at one node, the intermediate tensor immediately before this node should be stored, and this tensor can be deleted after the intermediate tensors immediately before its child nodes are all generated. 
For a tree traversal problem, two basic strategies, namely the deep first traversal (DFT) and the breadth first traversal (BFT) are often used. In BFT, 
one needs to store the intermediate tensors layer by layer, and the intermediate tensors for the current layer can only be deleted after those intermediate tensors for the next layer have been generated, as a result the memory cost in this approach is
\begin{align}
    \memcost^{\BFT} = \max_{l}\left(w_l2^{d_l}\right), 
\end{align}
where $d_l$ denotes to the rank of immediate tensors immediately before the $l$-th layer.
In DFT, only the intermediate tensors along one path from the root to a leaf need to be stored. Moreover, the intermediate tensors in the twig do not need to be stored since there will be no branching. As a result the memory cost in this approach is
\begin{align}
    \memcost^{\DFT} = \sum_{l \in {\rm branch}}2^{d_l}.
\end{align}
For computing uncorrelated amplitudes, $\memcost^{\BFT}$ scales linearly as $k$ since $w_l$ generally quickly approaches $k$. Therefore for large $k$ $\memcost^{\BFT}$ would soon explode, and DFT is the method of choice in this case. The memory cost of these two strategies are demonstrated in Fig.~\ref{fig:traversal}.

To this end we discuss several important differences between our method and the method used in Ref.~\cite{KalachevYung2021} (both aim for reducing the linear scaling of the multi-amplitude cost against the cost of computing a single amplitude), for example: 1) The reusable intermediate tensors are known before hand for us, which are simply the intermediate tensors immediately before each node in the tree, while in Ref.~\cite{KalachevYung2021} the reusable intermediate tensors are dynamically identified and then saved into a global cache (we do not need to use a dynamical global cache); 2) The memory usage and the computational cost are known before hand in our method for a given TNCO, while in Ref.~\cite{KalachevYung2021} they can only be known after the computation is done; 3) The tensor contraction patterns, including the order and the sizes of the tensors being contracted, are all known before hand for us. Due to these features of our method, we can systematically achieve the optimal multi-amplitude computational cost while at the same time maintaining the lowest memory cost. Moreover since the tree as well as the tensor contraction pattern are completely static, we can apply very fine-grained optimizations on multi-core architectures to increase the compute density and accelerate the computation.




\section{Reducing the cost of ma-TNC}

\subsection{Searching for better TNCO for ma-TNC}
\begin{figure}[htbp]
    \centerline{\includegraphics[width=0.6\textwidth]{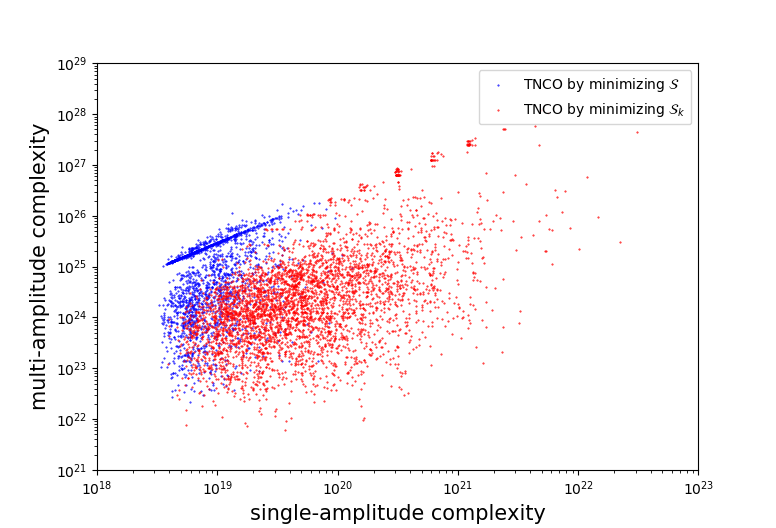}}
    \caption{Scatter plot of the single-amplitude and multi-amplitude computational complexities for $2000$ TNCOs found by minimizing the single-amplitude loss function in Eq.(\ref{eq:single_loss}) (blue dot) and for $4000$ TNCOs by minimizing the multi-amplitude loss function in Eq.(\ref{eq:full_loss}) (red dot).}
    \label{fig:paths}
\end{figure}

Since for a given TNCO we can compute the single-amplitude computational cost
\begin{align}\label{eq:single_loss}
\singlecpx = \sum_{l=0}^n\singlecpx_{l+1:l},
\end{align}
and the multi-amplitude computational cost
\begin{align}\label{eq:full_loss}
\fullcpx = \sum_{l=0}^{n} w_{l+1} \singlecpx_{l+1:l},
\end{align}
as shown in the main text, we can use them as the loss functions to minimize when we search for a near-optimal TNCO. The effectiveness of using the multi-amplitude loss function in Eq.(\ref{eq:full_loss}) compared to the existing approaches in which the cost of a single amplitude in Eq.(\ref{eq:single_loss}) (or a correlated batch) is minimized is demonstrated in Fig.~\ref{fig:paths}. We can clearly see that generally using the single-amplitude loss function one could get a lower single-amplitude cost but a higher multi-amplitude cost, while using the multi-amplitude loss function one could get a higher single-amplitude cost but a lower multi-amplitude cost. Since for simulating RQCs a number of uncorrelated amplitudes are required, minimizing the multi-amplitude loss function is the method of choice.



\subsection{The first-level slicing}\label{sec:sliceoverhead}

\begin{figure}[htbp]
    \centerline{\includegraphics[width=0.8\textwidth]{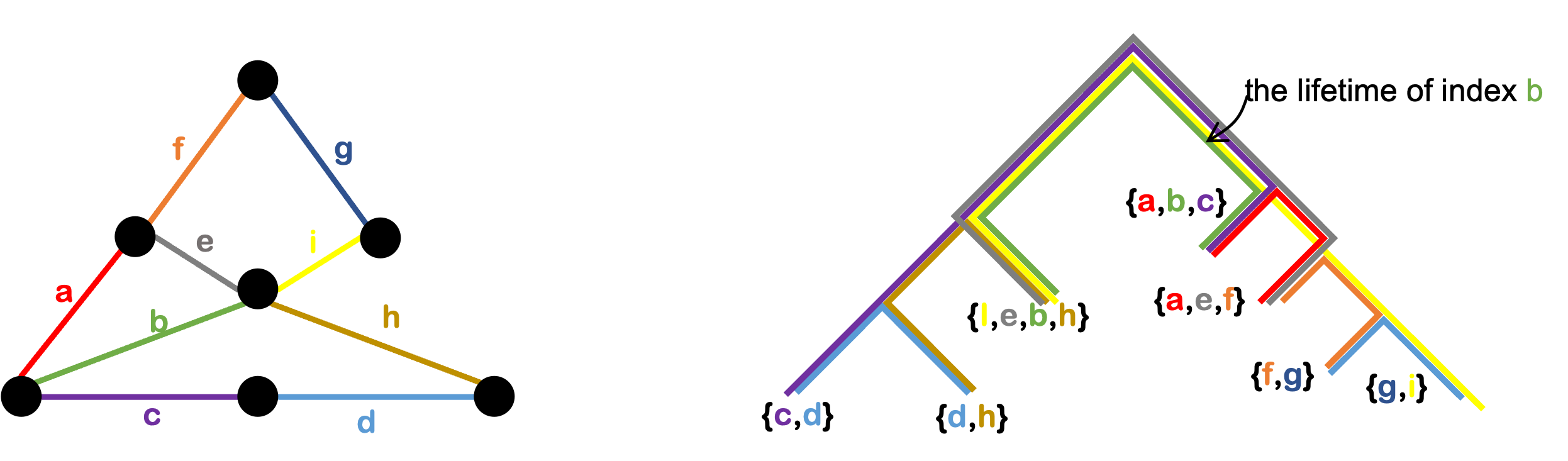}}
    \caption{Left: an example tensor network with $7$ tensors. Right: the contraction tree for contracting the tensor network on the left, which shows the order of contracting the tensor network (a different concept from the reuse tree in Fig.~\ref{fig:tree}). On the contraction tree, each tensor is represented as the collection of its indices. Each index is represented as a colored line and the life time of it is simply the length of the line. The pair-wise tensor contraction is performed from down to top whenever there is an intersection between two tensors. An index is contracted if two edges with the same color which both correspond to this index joins on the contraction tree. As indicated in the contraction tree, the indices in the dense region at the top can be sliced with little overhead. }
    \label{fig:lifetime}
\end{figure}

Slicing is an indispensable technique when applying the TNC algorithm for large TN, which transforms the original TN into the sum of many smaller TNs (slices) by slicing a number of indices of the original TN, such that the largest intermediate tensor during the contraction of each smaller TN could be easily handled (on a single CPU or GPU). Slicing also provides a natural way for perfect parallelization among different slices since there will be no data communication for contracting those slices. One minor drawback of slicing is that it would generally result in computational overhead. Nevertheless, by properly choosing the indices to be sliced, it is possible to achieve a very moderate computational overhead~\cite{HuangChen2021}.

In our previous work~\cite{ChenYang2022} we have proposed a lifetime based method to understand the computational overhead resulting from slicing, as well as to systematically reduce the overhead by finding a better set of indices to slice over. Here we will not go into the details of the lifetime theory but will briefly review the main results of it: 1) Each index in the TN has a \textit{lifetime}, which begins when the tensors that the index lives on are being contracted for the first time, and ends when this index is contracted; 2) Slicing will induce no computational overhead if and only if the lifetimes of the sliced indices cover the whole TNCO (\textit{the ideal setting}); 3) To reach the smallest overhead, the lifetimes of the sliced indices should cover the computation-intensive region as much as possible. Based on those facts, a \textit{slice refiner} was proposed which finds an optimal set of sliced indices to reduce the computational overhead for computing a single amplitude. The lifetime based method is demonstrated in Fig.~\ref{fig:lifetime}, from which we can see that the indices in the dense regions at the top of the contraction tree can be sliced with little overhead.

As is shown in the main text, the computation-intensive regions and the memory-intensive regions overlap in sa-TNC, but are separated in ma-TNC. In the first-level slicing, a major goal is to reduce the memory cost, therefore for lower memory cost one should generally slice the indices in the memory-intensive region. However, for ma-TNC the computation-intensive region happens later than the memory-intensive region, and for lower computational cost one should slice the indices in the computation-intensive region. As a result there is a conflict for slicing in our ma-TNC: it is difficult to balance the memory cost and the computational cost at the same time as in the sa-TNC. Since memory is a hard requirement, the slicing overhead of ma-TNC will be much larger than sa-TNC in general. Nevertheless, we generalize the slice refiner to work with the multi-amplitude loss function, which tends to choose the sliced indices that cover the tail part as much as possible under the same memory constrain.

\begin{figure}[htbp]
    \centerline{\includegraphics[width=0.6\textwidth]{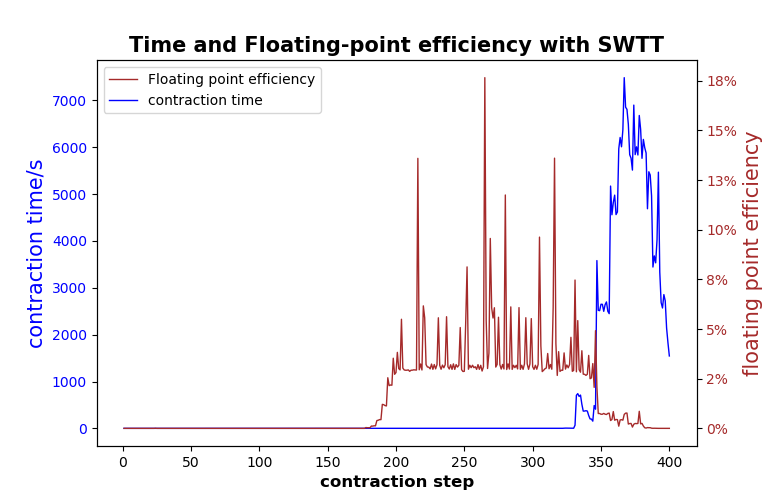}}
    \caption{Distribution of the pair-wise tensor contraction time (left axis) and Flops efficiency (right axis) using our ma-TNC for computing $3M$ amplitudes of Sycamore RQC with depth $20$, where the pair-wise tensor contractions are performed one by one along the tensor network contraction order and swTT is used for each tensor contraction. }
    \label{fig:distribution}
\end{figure}

\section{Fused tensor network contraction algorithm}

In the previous sections we have mostly focused on reducing the theoretical memory cost and the computational cost of our ma-TNC. In practical implementation, we meet the challenge of extremely low compute density for our ma-TNC algorithm, which is another consequence from the separation of the memory-intensive region and the computational-intensive region.

In Fig.~\ref{fig:distribution} we plot the distribution of the floating-point efficiency and the runtime for each pair-wise tensor contraction when using swTT together with our ma-TNC to compute $3M$ amplitudes of Sycamore. We can see that the last $50$ steps accounts for more than $95\%$ of the time cost, but the Flops efficiency of these steps are extremely low. Therefore performing tensor contractions one by one independently on the new Sunway supercomputer would at most have a single-precision efficiency of $8$ Pflops (this behavior is likely to be very similar for other architectures), which is in comparison with sa-TNC for which a $308.6$ Pflops single-precision efficiency has been achieved~\cite{ChenYang2022}.

\begin{figure}
\center
\includegraphics[width=0.5\columnwidth]{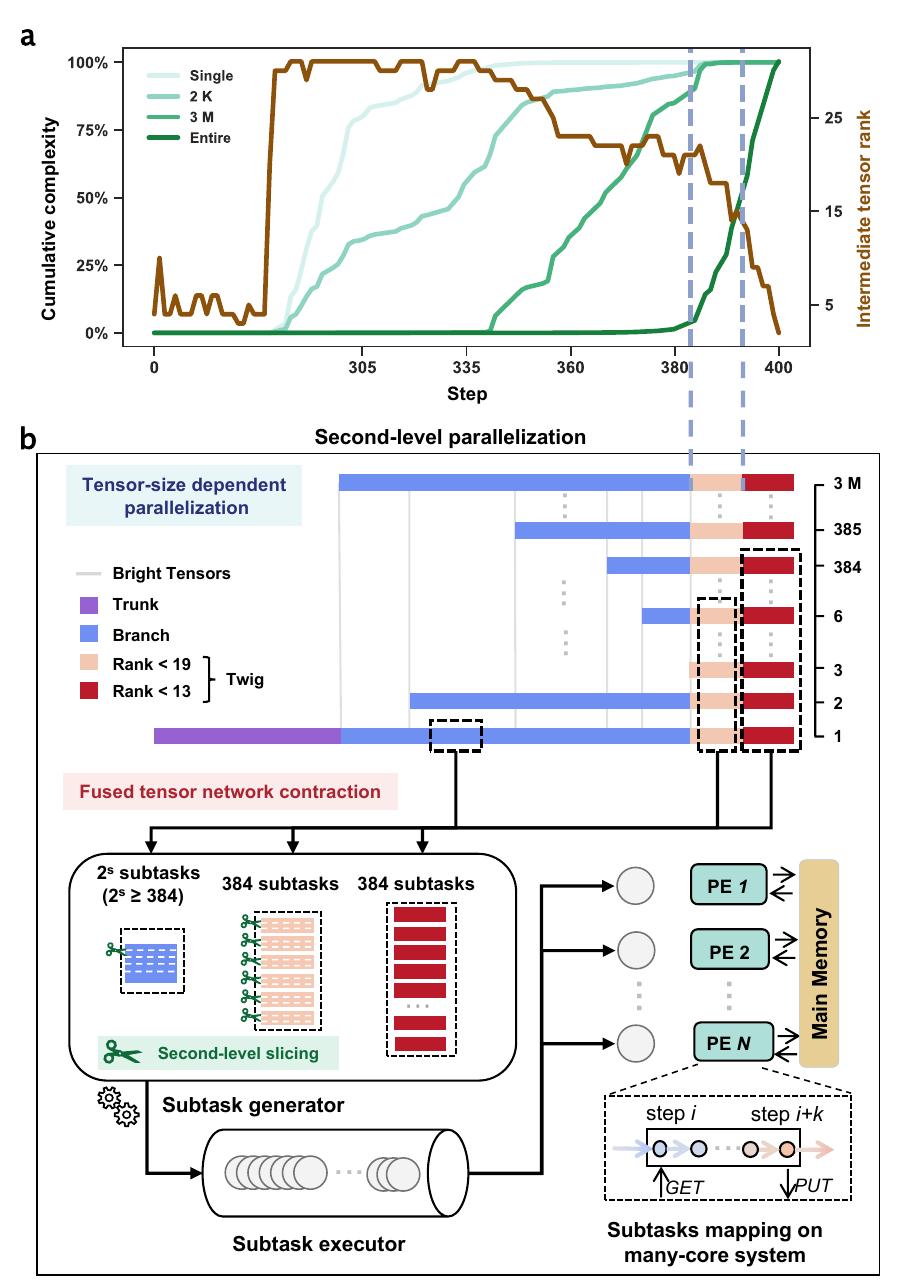}
\caption{(a) Distribution of the cumulative computational cost (the blue lines to the left axis) and the intermediate tensor rank (the brown line to the right axis) along the tensor network contraction order. (b) Second-level parallelization on one SW26010P CPU for computing one slice (from the first-level slicing). In the trunk swTT is used for tensor contraction. In the branch and twig, 
the fused TNC algorithm is used which performs several tensor contractions locally on each core before fetching the result into the main memory. A tensor-size dependent parallelization scheme is also used to dynamically adjust the parallelization strategies for different tensor sizes such that the cores on a CPU can be fully utilized.
}
\label{fig:fusedTNCoverview}
\end{figure}

From Fig.~\ref{fig:distribution} we can see that the runtime is dominated by the calculations in the end with very low computational efficiency. 
To understand the drastic drop of computational efficiency, we plot the theoretical computational cost as functions of the contraction step for sa-TNC and our ma-TNC with different $k$s, as shown in Fig.~\ref{fig:fusedTNCoverview}(a). We can see that for the case of computing one amplitude, most computation falls into the region of large intermediate tensors (with tensor ranks $\approx 30$). In contrast, for the case of computing millions of uncorrelated amplitudes, most computation concentrates in the branch where the tensors are much smaller (tensor ranks $\approx 20$). The smaller sizes of the tensors involved in the ma-TNC algorithm result in a significant decrease of the compute density, and a lower computational efficiency, compared to the sa-TNC. 
In the extreme case of computing all the amplitudes of Sycamore, the last $5\%$ steps where the tensor sizes are smaller than $2^{20}$ account for more than $90\%$ of the computational cost. If one performs the tensor contractions step by step as in the standard approach, the drop of computational efficiency will largely hinder the gain of a lower computational cost. 

Our solution to the low compute density of ma-TNC is a fused TNC algorithm, the main idea of which is to perform several successive tensor contractions together to reduce the data movement between the main CPU memory and the local memories of the cores. 
In practice, the fused TNC algorithm is complemented with a tensor-size dependent parallelization scheme to fully utilize all the cores: for large tensors that would produce more slices than the cores to feed, we simply parallelize over the slices; for small tensors which can be directly stored on the local memory of a single core, we parallelize over different amplitudes; for medium-size tensors that are not able to produce enough slices, we parallelize over both the slices and the amplitudes. The overview of the fused TNC algorithm and the adaptive parallelization scheme are demonstrated in Fig.~\ref{fig:fusedTNCoverview}(b).

The detailed explanations of the algorithm is presented in the following, which is based on the SW26010P CPU but the idea could be straightforwardly generalized to other heterogeneous architectures by adjusting the hyperparameters of the algorithm. 

\subsection{Fusing successive tensor contractions together}

\begin{figure}[htbp]
    \centerline{\includegraphics[width=0.6\textwidth]{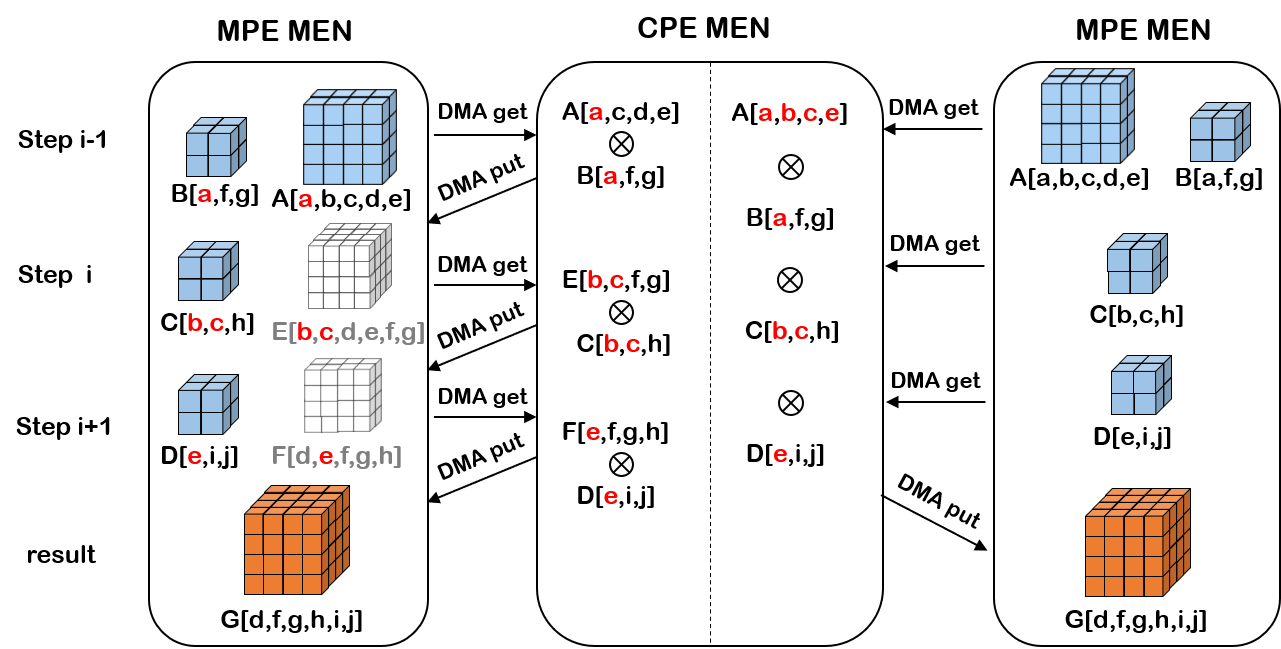}}
    \caption{A schematic demonstration of our fused tensor network contraction algorithm which performs the tensor contractions $G[d,h,l,j]=A[a,b,c,e]\times B[a,f,g]\times C[b,c,h]\times D[e,l,j]$ locally on each core of one CPU, where the tensors $A, B, C, D$ are slices of larger tensors in the original tensor network. The result tensor $G$ is fetched back to the main memory of the CPU in the end.}
    \label{fig:fusedTNC}
\end{figure}

A standard approach for TNC is to perform pair-wise tensor contractions, which is used in previous works. On a heterogeneous architecture, the process of a single tensor contraction can be abstracted into three steps: 
\emph{copy-in, computation (and communication), copy-out}. For contraction of two large tensors, the cost of the computation step is large compared to the copy-in and copy-out steps, and in this case the overhead of the data movement can be neglected. However, for the TN formed in the TNC algorithm, contractions typically happen between a high-rank tensor (rank $\approx 30$) and a low-rank tensor (rank $\leq 6$), for which the compute density is at the same level as the data movement. As a result, data exchange consumes most of the runtime for the TNC algorithm on most bandwidth-constrained architectures. For our ma-TNC, the tensors involved in the tensor contractions are even smaller (the higher rank tensor has a rank $\leq 20$).

Generally, data exchange often comes from the lacking of data locality. 
For a pair-wise tensor contraction $C=A\times B$, if the output tensor $C$ is just one of the input tensors of the next step, it is possible to avoid the copy-out step of the tensor $C$, which forms the basic idea of our fused TNC algorithm. 
The main idea of our fused TNC algorithm for several successive tensor contractions is demonstrated in Fig.~\ref{fig:fusedTNC}.

In practice, there are three situations which prevent our fused TNC algorithm from achieving ideal speedup: 1) $C$ is not used for the next contraction; 2) $C$ is too large to be stored in the local memories of the cores; 3) if a bright tensor is met one has to stop the fusing since the intermediate tensor here needs to be explicitly stored and reused.
Those situations as well as the solutions to them are shown in detail in the following.

\subsection{Rearranging the TNCO}

\begin{figure}[htbp]
    \centerline{\includegraphics[width=0.8\textwidth]{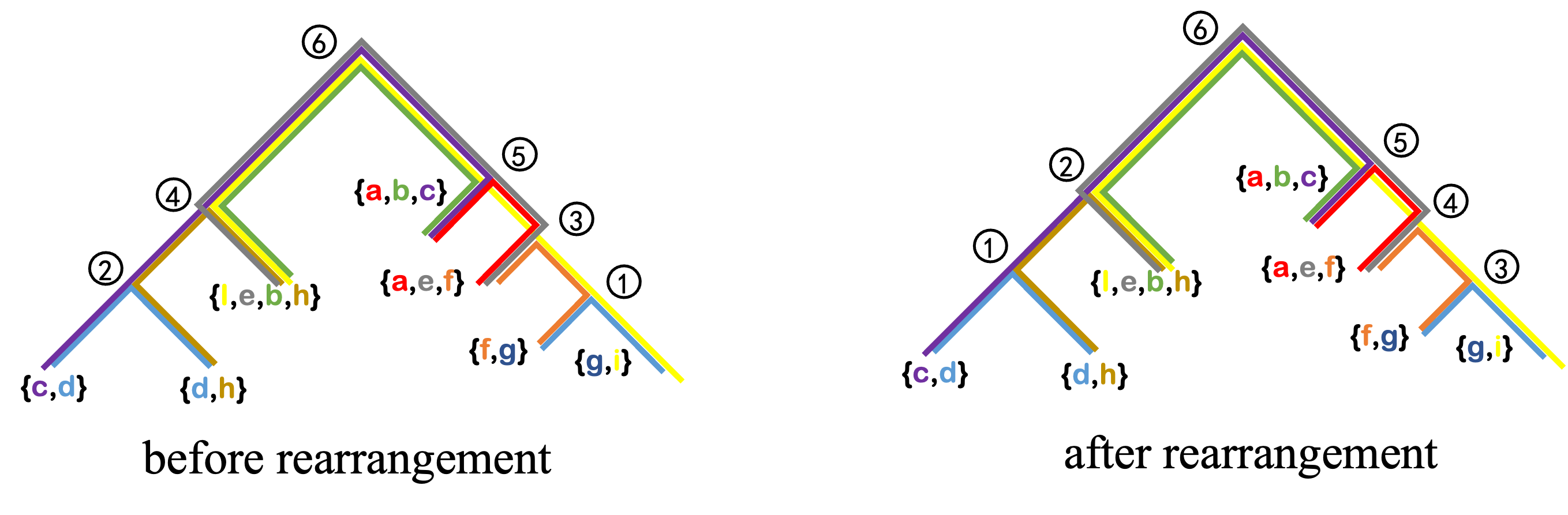}}
    \caption{Rearranging the TNCO for higher data dependency between contraction steps. The number at one intersection point (where a tensor contraction is performed) indicates the order of this tensor contraction in the whole TNCO. The two contractions trees on the left and right are completely equivalent in terms of computational cost. However, the contraction tree on the right has higher data dependency and is better suited for our fused TNC algorithm.}
    \label{fig:rearrange}
\end{figure}

The first situation could be overcome by proper rearrangement of the TNCO, without affecting the time cost.
Given a contraction tree for a (sub-)TN, there could be many equivalent TNCOs with the same computational cost. 
This feature can be utilized to rearrange the TNCO to increase the data dependency between successive tensor contraction steps, which is demonstrated in Fig.~\ref{fig:rearrange}.

\subsection{Lifetime based second-level slicing}\label{sec:lifetime}

\begin{figure}[htbp]
    \centerline{\includegraphics[width=0.5\textwidth]{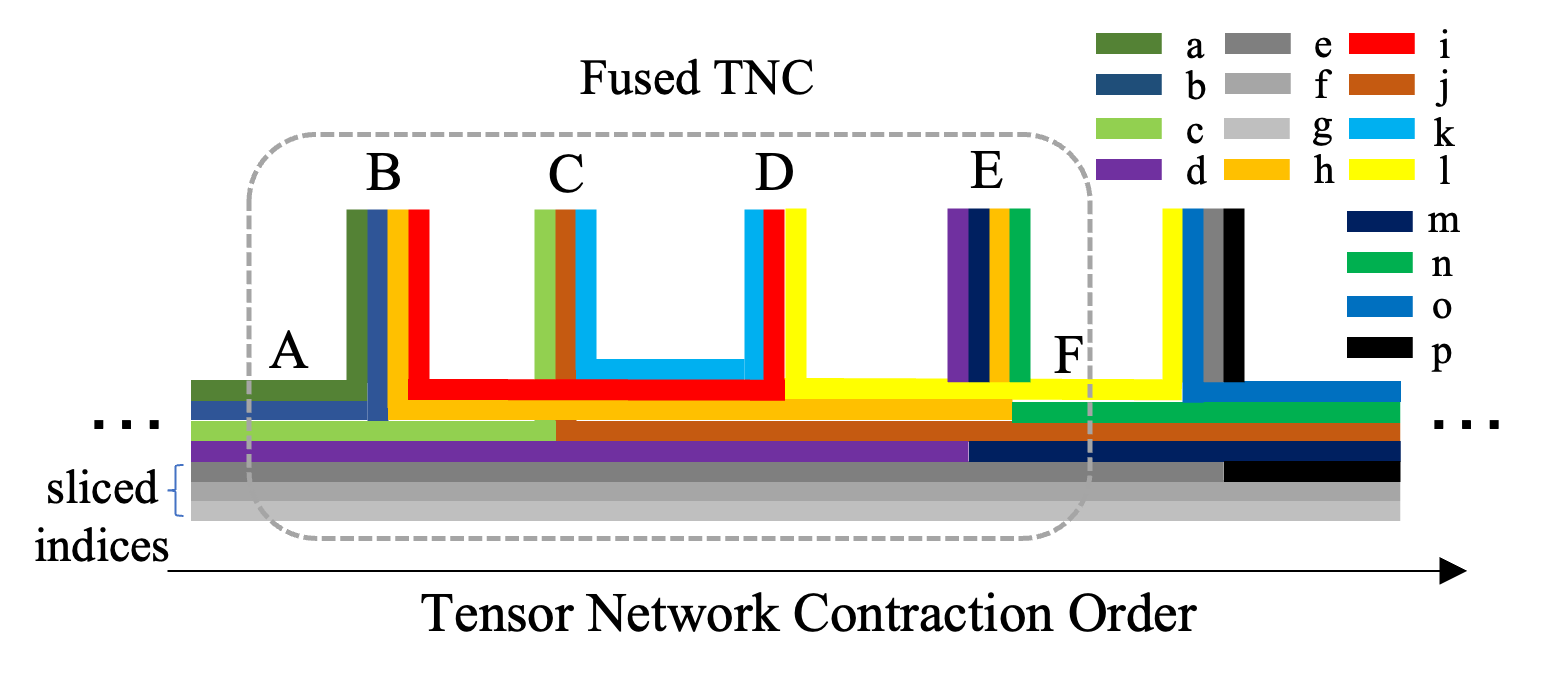}}
    \caption{Demonstration of the lifetime based second-level slicing under a threshold $d^t=4$. The tensor contractions $F[e,f,g,j,l,m,n] = A[a,b,c,d,\textbf{\textit{e, f, g}}] \times B[a,b,h,i] \times C[c, j, k] \times D[i,k,l] \times E[d,h,m,n]$ are fused together and then the result tensor $F$ is stacked such that there will be no computational overhead induced by slicing. The gray lines, \emph{i.e.} $\textbf{\textit{e, f, g}}$, are the sliced indices.}
    \label{fig:lifetime-slice}
\end{figure}

The situation is more complicated if the intermediate tensors are too large such that they do not fit into the local memory. Interestingly, this situation is very similar to the situation we are faced with in the first-level slicing: one would like to slice a large TN into smaller TNs such that each smaller TN fits into the ``local computational unit''. In the first-level slicing the local computational unit is the CPU while in the second-level slicing the local computational unit is the core. However, there is also an important difference: in the second-level slicing the tensors are shared by all the cores while in the first-level slicing each CPU takes a slice and no data communication is allowed except when the computation on each CPU has been finished. Due to this feature of the second-level slicing, we could perform the sliced computation without any computational overhead compared to the non-sliced one, which is shown in the following.


First of all, we do not want to directly merge the second-level slicing into the first-level slicing, since they happen in very different regions of the TNCO and rarely overlap. According to the lifetime based theory, the overhead introduced by this merging is likely to grow exponentially with the number of new indices been sliced. Therefore we focus on each slice resulting from the first-level slicing and perform a second-level slicing locally on each CPU on top of it.



The key point of the lifetime based second-level slicing strategy without any computational overhead is to do \textit{stacking} immediately before the lifetimes of any one of the sliced indices end.
Stacking is the inverse operation of slicing. Slicing an index drops a dimension of each involved tensor and produces two smaller tensors, while stacking join the two smaller tensors into a larger one with one more dimension. 
If we stack the sliced indices immediately after several successive tensor contractions where all those sliced indices are involved, those sliced indices will cover the whole sub-TNCO and by stacking we restore the original TN at the end of this sub-TNCO. As such we effectively achieve the ideal setting of the lifetime based slicing which has no computational overhead. In practice, starting from an intermediate tensor in the TN, we can count the highest rank of the tensors along the TNCO, as well as the number of indices of each intermediate tensor generated in successive tensor contractions that participate in the contraction, and if one of them is larger than a threshold $d^t$ or if a bright tensor is met, we stop fusing. Our lifetime based second-level slicing is demonstrated in Fig.~\ref{fig:lifetime-slice}.

\subsection{Exploring the sparsity of the reuse tree}

The fusing of tensor contractions has to stop once a branching bright node is met, since otherwise the computational cost will increase. This fact also limits the number of tensor contractions that can be fused together. Nevertheless, by dynamically exploring the sparsity of the reuse tree, this problem can be largely overcome. The sparsity of the reuse tree is defined as the ratio between the width of the previous layer and the current layer, which is
\begin{align}
\sparsity_l = \frac{w_{l-1}}{w_{l}},
\end{align}
which satisfies $0.5\leq \sparsity_l\leq 1$ and describes the density of branching between the $l-1$-th and $l$-th layers. $\sparsity_l\approx 1$ means almost no branching (the twig) and $\sparsity_l\approx 0.5$ means that there is a branching at every node in the $l-1$-th layer (the branch). 

\begin{table}[!htb]
  \begin{center}
    \caption{Growth of the width of each layer in the reuse tree against the tensor contraction step for Sycamore.}
    \label{tab:wl}
    \begin{tabular}{ccc}
    \hline
    \hline 
    Position of bright tensor & Number of output indices  & $w_l$  \\
    $306$ & $3$ & $8$  \\
    $317$ & $4$ & $16$  \\
    $323$ & $5$ & $32$  \\
    $331$ & $10$ & $1024$  \\
    $332$ & $14$ & $16384$  \\
    $345$ & $16$ & $65536$  \\
    $347$ & $20$ & $981285$  \\
    $357$ & $22$ & $2123441$  \\
    $363$ & $24$ & $2737958$  \\
    $370$ & $26$ & $2931054$  \\
    $371$ & $27$ & $2965352$  \\
    $373$ & $29$ & $2991260$  \\
    $379$ & $30$ & $2995698$  \\
    $381$ & $31$ & $2997840$  \\
    $382$ & $33$ & $2999468$  \\
    $383$ & $34$ & $2999722$  \\
  \hline
    \end{tabular}
  \end{center}
\end{table}

\begin{figure}[htbp]
    \centerline{\includegraphics[width=0.6\textwidth]{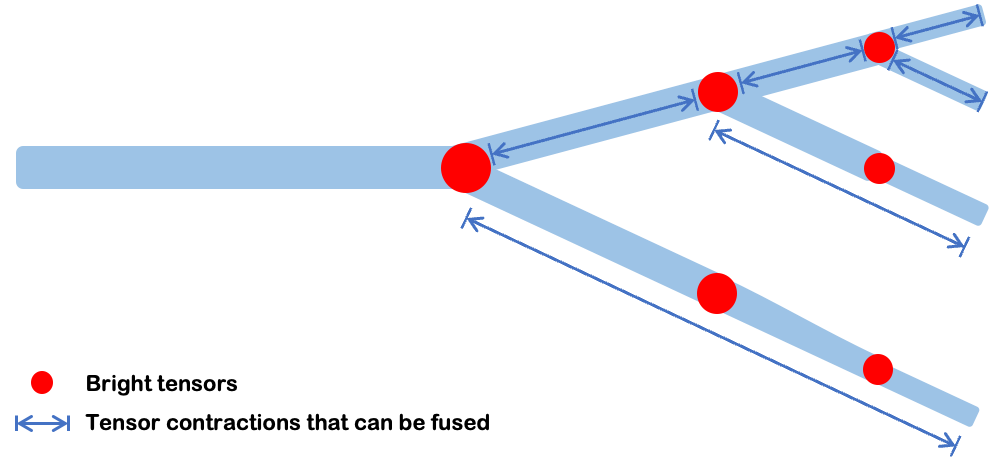}}
    \caption{Dynamically extending the range of the fused tensor network contraction over the bright nodes without branching. }
    \label{fig:dynamic}
\end{figure}

In Table.~\ref{tab:wl} we show the growth of $w_l$ as $l$ increases, where we can see that $w_l$ approximately grows exponentially for the first few bright tensors and then keeps approximately constant, which means that $\sparsity_l\approx 0.5$ for small $l$ and $\sparsity_l\approx 1$ afterwards (therefore the reuse tree is very sparse).
The sparsity of the reuse tree allows us to significantly extend the range of the fused TNC algorithm, as demonstrated in Fig.~\ref{fig:dynamic}. Whenever there is no branching at some bright tensor, our fused TNC algorithm can simply fuse across this tensor as for the rest tensors. By exploring this fact, we are able to increase our computational efficiency by three times.

In complementary to the fused TNC algorithm, a tensor-size dependent parallelization scheme is used to fully utilize all the cores. On our SW26010P CPU (see Sec.~\ref{sec:cpu} for details), this is done as follows. If the tensor sizes are too large that the total local memories are not enough to store them, then during the lifetime based second-level slicing we will slice the tensors to produce slices with size $2^{12}$ each (so as to fit into the local memory), and send each slice to one core. If the number of slices produced is larger than the number of cores (namely $384$), we parallelized over slices. If the number of slices produced is smaller than $384$, we take the computations for computing a few amplitudes simultaneously such that the number of slices times the number of amplitudes is equal to or larger than $384$, and we parallelize over both slices and amplitudes. If the tensors are smaller than (include) $2^{12}$ such that they can be simply stored on the local memory of a single core, we simply parallelize over $384$ amplitudes.
In addition, if tensor contractions can not be fused due to the lifetime theory (which usually happens for very larger tensors in the trunk), we will use swTT instead.

\section{Performance}
\subsection{Platform}\label{sec:cpu}

\begin{figure}[htbp]
    \centerline{\includegraphics[width=0.5\textwidth]{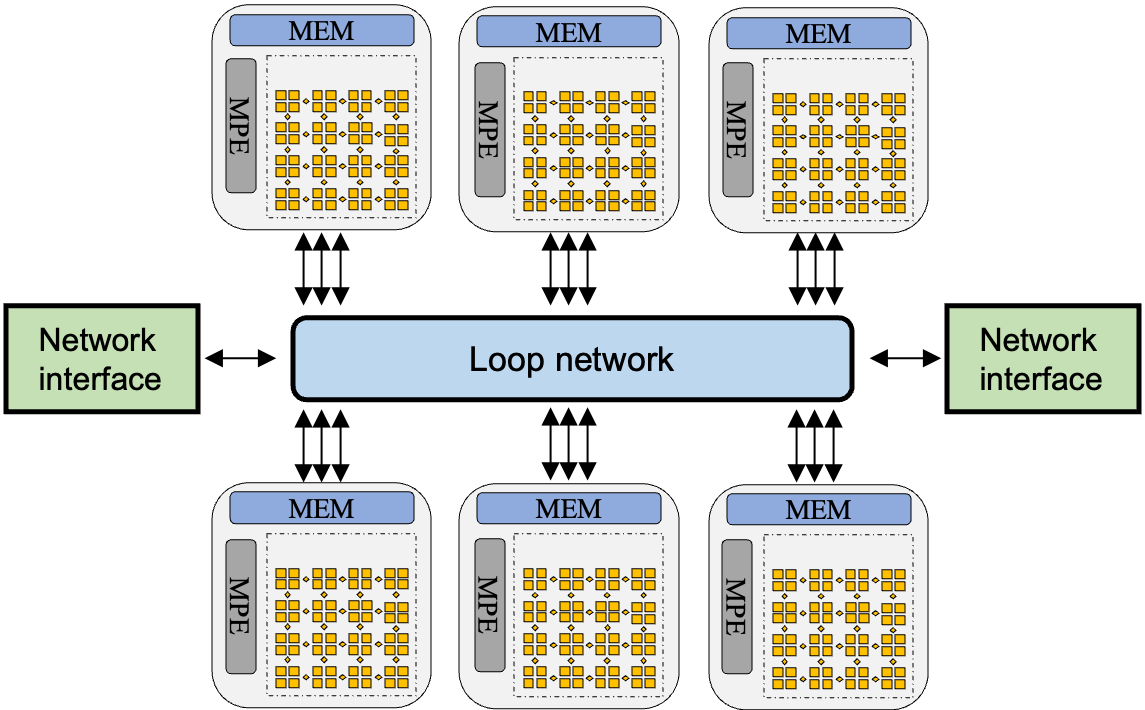}}
    \caption{A schematic diagram of the SW26010P CPU.}
    \label{fig:CPU}
\end{figure}

The building block of the new Sunway supercomputer is the SW26010P CPU, the architecture of which is shown in Fig.~\ref{fig:CPU}.
On each CPU, there are $6$ core groups (CGs), each of which contains $64$ computation process elements (CPE). The loop network connects the CGs and provides communication access among them. In each CG, there is a $16$ GB main memory and a management process element (MPE) for thread management. Process level parallelization are often performed between CPUs, or between CGs of one CPU. In this work, we used the whole CPU as one process such that we have a total of $96GB$ memory on each process.
CPEs perform the actual computation. On each CPE, there is a $256$ KB local memory (LDM) and $32$ $512$-$bit$ vectorized registers. Direct memory access (DMA) is applied for data exchange between LDM and the main memory, with a bandwidth of $300$ GB/s on a CPU. The single-precision performance of a CPU is $14$ Tflops. Thread level parallelization are between CPEs.

\subsection{Performance and Efficiency}
\begin{figure}[htbp]
    \centerline{\includegraphics[width=0.5\textwidth]{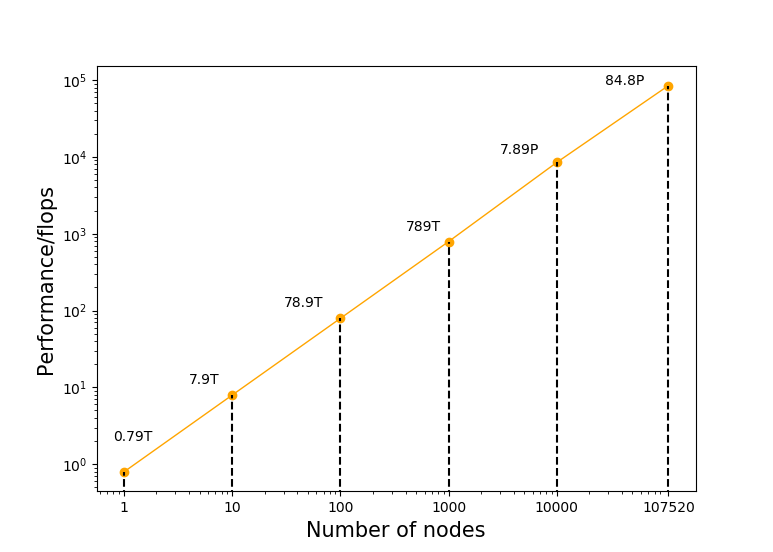}}
    \caption{The strong scaling for computing $3M$ uncorrelated amplitudes for the Sycamore random quantum circuit with depth $20$ using our ma-TNC.}
    \label{fig:scaling}
\end{figure}
In our simulation of calculation $3M$ uncorrelated amplitudes, we achieve a sustaining performance of $84.8$ Pflops using $107,520$ SW26010P CPUs (nodes). Due to the independent subtasks generated by the first-level slicing, the performance scaling with respect to the number of CPUs is approximately linear, as shown in Fig.~\ref{fig:scaling}.

\begin{figure}[htbp]
    \centerline{\includegraphics[width=0.5\textwidth, height=0.35\textwidth]{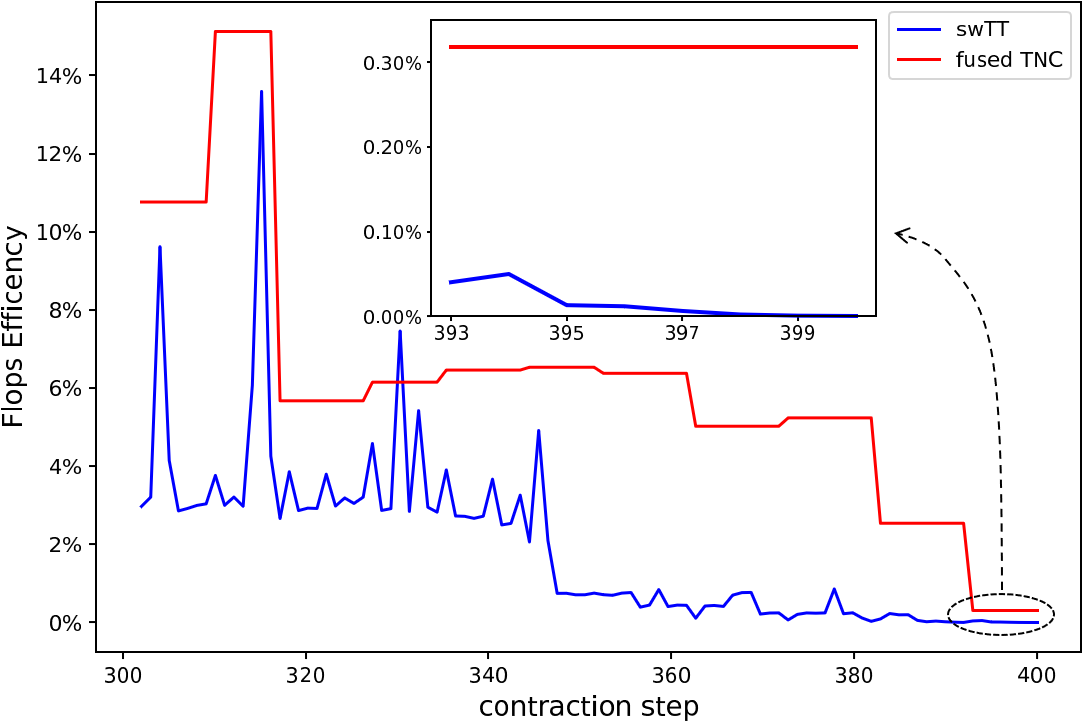}}
    \caption{Computational efficiency of ma-TNC using swTT and using our fused tensor network contraction. The efficiency of swTT is calculated step by step, and for fused tensor network contraction, it is calculated for all the successive contraction steps that are fused together.}
    \label{fig:efficiency}
\end{figure}

Using our fused TNC algorithm assisted by the adaptive parallelization scheme, the computational efficiency of our ma-TNC has been greatly improved. The comparison between the computational efficiency after using those optimization and the computational efficiency using swTT only is shown in Fig.~\ref{fig:efficiency}, from which we can clearly see the dramatic increase in the computational efficiency (especially in the branch and twig).

\bibliographystyle{apsrev4-1}
%